\documentclass[runningheads]{llncs}
\usepackage{graphicx}
\usepackage{paralist}
\usepackage{microtype}
\usepackage{units}
\usepackage{color}
\usepackage{hyperref}

\begin{document}
\title{Music Instrument Classification Reprogrammed}
%
%
\author{Hsin-Hung Chen\orcidID{0000-0002-3406-741X} \and
Alexander Lerch\orcidID{0000-0001-6319-578X} }
%
\authorrunning{H.\ Chen \and A.\ Lerch}
%
\institute{Music Informatics Group, Georgia Institute of Technology}
%
\maketitle              
\begin{abstract}
The performance of approaches to Music Instrument Classification, a popular task in Music Information Retrieval, is often impacted and limited by the lack of availability of annotated data for training. We propose to address this issue with ``reprogramming,'' a technique that utilizes pre-trained deep and complex neural networks originally targeting a different task by modifying and mapping both the input and output of the pre-trained model. We demonstrate that reprogramming can effectively leverage the power of the representation learned for a different task and that the resulting reprogrammed system can perform on par or even outperform state-of-the-art systems at a fraction of training parameters. Our results, therefore, indicate that reprogramming is a promising technique potentially applicable to other tasks impeded by data scarcity.

\end{abstract}
\section{Introduction}

The task of Music Instrument Classification (MIC) aims at automatically recognizing the musical instruments playing in a music recording. MIC can provide important information for a variety of applications such as music recommendation, music discovery, and automatic mixing. In recent years, Deep Learning (DL) models have shown superior performance in practically all Music Information Retrieval (MIR) tasks including MIC. However, the lack of large-scale annotated data remains a major problem for data-hungry supervised machine learning algorithms in this field \cite{Defferrard2017FMAAD,humphrey2018openmic,DBLP:conf/ismir/GururaniSL19}.  Beyond small-scale expert-annotated datasets, larger datasets are often collected by crowd-sourcing annotations, which leads to noisy and sometimes incomplete labels. For example, the majority of labels in the OpenMIC dataset \cite{humphrey2018openmic}~---a popular dataset for polyphonic MIC---~are missing; this data scarcity can negatively impact the training of complex classifiers for this multi-label task.

\begin{figure}[t]
\vspace{-5mm}
\centering
\includegraphics[width=\textwidth]{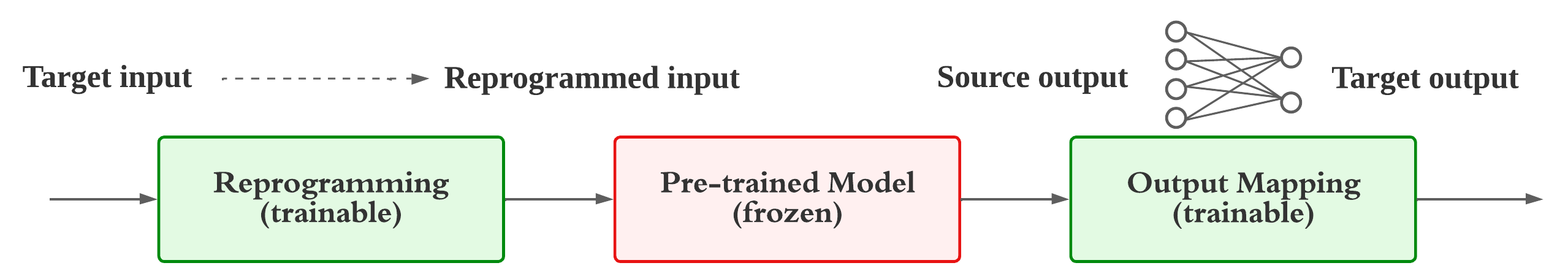}
\caption{The concept of model reprogramming. }
\label{fig:repr}
\vspace{-7mm}
\end{figure}

One established approach to address this data challenge is transfer learning. In this approach, the knowledge of a source domain with sufficient training data is transferred to a related but different target domain with insufficient training data. This knowledge transfer is often achieved by either directly using a pre-learned representation as classifier input or by fine-tuning a pre-trained source-domain model with the target-domain data. For instance, the VGGish representation \cite{hershey2017cnn}, trained on a wide variety of audio data, has been successfully utilized for MIC \cite{DBLP:conf/ismir/GururaniSL19}.
\emph{Model reprogramming} aims at expanding the transfer learning paradigm by treating the source-domain models as unmodified black-box machine learning models extended only by input pre-processing and output post-processing. Model reprogramming was first introduced in 2018 by Elsayed et al.\ \cite{elsayed2018adversarial}, who showed that a trainable universal input transformation function can reprogram a pre-trained ImageNet model (without changing the model weights) to solve the MNIST/CIFAR-10 image classification task with high accuracy. 
Figure~\ref{fig:repr} illustrates the basic concept of model reprogramming: a trainable model for input reprogramming modifies the input data to be fed into the frozen black-box model pre-trained on the source task, followed by an output transformation that maps the outputs of the pre-trained model to the target categories. 
Thus, the reprogramming layer serves to ``reprogram'' the pre-trained model to work with a new target task with different input data and different target classes. Since the complexity of the input and output transformation can be low, model reprogramming can combine the advantage of leveraging a powerful deep pre-trained representation with the advantages of 
\begin{inparaenum}[(i)]
    \item   reduced training complexity and 
    \item   reduced data requirements.
\end{inparaenum}
Reprogramming methods have been successfully applied to various tasks such as medical image classification \cite{tsai2020transfer}, time-series classification \cite{yang2021voice2series}, and language processing \cite{neekhara2018adversarial}. Results show that reprogramming methods can perform on par or better than state of the art methods, thus demonstrating the feasibility of reprogramming pre-trained models and showing the potential of this method for improving the performance on other tasks with low amounts of data. 

In this work, we investigate reprogramming for the task of MIC with the incompletely labeled OpenMIC dataset. We choose a pre-trained state-of-the-art audio classification model and extend it by input pre-processing and label-mapping. We provide extensive results on various input processing configurations. The main contributions of the paper are
\begin{compactenum}[(i)]
    \item   the presentation of a system with low complexity that is able to outperform state-of-the-art MIC systems, and
    \item   the introduction of the reprogramming paradigm to the field of MIR with a multitude of applications with insufficient training data.
\end{compactenum}   

The remainder of the paper is structured as follows. The following Sect.~\ref{sec:relate} presents a brief overview of relevant work. The pre-trained model and the proposed reprogramming methods are introduced in Sect.~\ref{sec:method}. The evaluation and analysis are presented in Sect.~\ref{sec:eval}. We conclude with a brief summary and directions of future work.

\section{Related Work}
\label{sec:relate}
This related work is structured into three main parts, an overview of MIC, a short survey of transfer learning, and recent work on reprogramming.

While earlier research on MIC has focused on the detection of instruments from audio only containing one instrument \cite{1661252,1661401,859069,Lostanlen2016DeepCN} or on the detection of the pre-dominant instrument in a mixture \cite{bosch2012comparison}, current research has focused on recognizing instruments in polyphonic and poly-timbral audio recordings containing multiple instruments playing multiple voices simultaneously.
Similar to other audio classification tasks, earlier systems tended to use traditional machine learning approaches with low-level audio features at the input \cite{859069,8256990} while modern approaches are dominated by neural networks. Li et al.\ \cite{Li2015AutomaticIR} proposed to learn features  for instrument recognition with a Convolutional Neural Network (CNN) using the MedleyDB dataset \cite{Bittner2014MedleyDBAM}. Hung et al.\ proposed to detect instrument activity at a high time-resolution and showed the advantage of pitch-conditioning on instrument recognition performance \cite{DBLP:conf/ismir/HungY18}. Gururani et al.\ \cite{DBLP:conf/ismir/GururaniSL19} introduced training attention-based models to the task for enhanced accuracy and implemented a partial binary cross-entropy loss to ignore missing labels in the OpenMIC dataset \cite{humphrey2018openmic}. Gururani and Lerch showed that a semi-supervised approach based on consistency loss to adapt and leverage the data with missing labels outperforms other systems \cite{9666061}. Despite previous efforts on data curation and annotation, the access to fully annotated data on a large scale remains a challenge.
The IRMAS dataset \cite{bosch2012comparison} is a polyphonic dataset with mixed genres, however, it targets predominant instrument recognition and is therefore not suitable for multi-instrument classification. The MedleyDB \cite{Bittner2014MedleyDBAM} and Mixing Secrets \cite{gururani2017mixing} datasets are both multi-track datasets with strong annotations of instrument activity. However, only a few hundred of songs are available, creating potential issues not only with respect to the dataset size itself but also regarding data distribution and diversity. The OpenMIC dataset for polyphonic instrument recognition, published by Humphrey et al.\ \cite{humphrey2018openmic}, presents a reasonably large sample size across various genres. Unfortunately, a considerable number of labels are missing as not all clips are labeled with all 20 instruments due to the crowd-sourced annotation process. Slakh2100 is another large music dataset containing mixed tracks of 34 instrument categories and with perfect annotation. However, it is synthesized and rendered from MIDI files instead of real recordings.
Transfer learning is an important tool in machine learning when facing the fundamental problem of insufficient training data. It aims to transfer the knowledge from a source domain to the target domain by relaxing the assumption that the training data and the test data must be independent and identically distributed \cite{tan2018survey}. It is based on the idea that a powerful representation, learned for a source task with large datasets, can be adapted to a related but not identical target task that lacks training data. 
For example, Qiuqiang et al.\ \cite{9229505} present the adaptation of pre-trained models such as ResNet \cite{he2016deep} and MobileNet \cite{Howard2017MobileNetsEC} to audio tagging tasks, showing the generalizability of systems pre-training on large-scale datasets to audio pattern recognition. Choi et al.\ \cite{Choi2017TransferLF} show that representations pre-trained on the music tagging task can be successfully transferred to various music classification tasks and can lead to competitive results. Jordi et al.\ introduced the ``musicnn'' representations, featuring a set of CNNs pre-trained on the music audio tagging task \cite{pons2019musicnn}. In the context of MIC, Gururani et al.\ \cite{DBLP:conf/ismir/GururaniSL19} successfully adopt the VGGish pre-trained representation  \cite{hershey2017cnn} as the input features for their attention-based model.

The promising results of transfer learning outlined above led to Tsai et al.\ framing two new research questions \cite{tsai2020transfer}:
\begin{inparaenum}[(i)]
    \item   ``is finetuning a pre-trained model necessary for learning a new task?'' and
    \item   ``can transfer learning be expanded where nothing but only the input-output model responses are observable?''
\end{inparaenum}
The attempt to answer these questions inspired by ideas of adversarial approaches led to the idea of ``reprogramming.''
Adversarial machine learning aims at manipulating the prediction of a well-trained deep learning model by designing and learning perturbations to the data inputs without changing the target model \cite{Biggio2018WildPT,elsayed2018adversarial,tsai2020transfer}. The success of these approaches shows that the classifier output can be changed just by modifying its input, and thus suggests that such methods might be applicable in a non-adversarial context by modifying the input of a pre-trained model to ``adapt'' the model to a target task. This leads to the concept of model reprogramming, also referred to as adversarial reprogramming \cite{elsayed2018adversarial}, which is a technique that aims at leveraging the knowledge of a model pre-trained for a different task by pre-processing the input and post-processing the output. Elsayed et al.\ showed that pre-trained ImageNet models can be reprogrammed to classifying MNIST and CIFAR-10 by adding learnable parameters around the input image \cite{elsayed2018adversarial}. Tsai et al.\ demonstrate the advantage of reprogramming on label-limited data such as biomedical image classification \cite{tsai2020transfer}, and combine reprogramming with multi-label mapping. Model reprogramming has also been used in tasks other than image classification such as natural language processing. For example, Hambardzumyan et al.\ propose Word-level Adversarial ReProgramming (WARP) for language understanding by adding learnable tokens to the input sequence \cite{hambardzumyan2021warp}. The evaluation shows that WARP outperforms all models with less parameters. Neekhara et al.\ demonstrate re-purposing character-level classification tasks to sentiment classification tasks by implementing a trainable adversarial sequence with surrounding input tokens \cite{neekhara2018adversarial}.  Yang et al.\ \cite{yang2021voice2series} applied reprogramming to acoustic models for time-series classification on the UCR Archive benchmark \cite{dau2019ucr}. The input audio is treated as time-series and is padded to be reprogrammed. The model achieves state-of-the-art accuracy on 20 out of 30 datasets with considerably fewer trainable parameters than the pre-trained models.

The presented reprogramming methods achieve promising results with simple learnable reprogramming operations. Therefore, we can conclude that reprogramming is an effective new transfer learning approach inspired by adversarial methods that could address data insufficiency problems by requiring less training complexity.

\section{Proposed Method}\label{sec:method}
%

\subsection{Pre-trained Model}
\label{sec:ast}

The criteria for choosing the pre-trained model to be used as black-box model in our reprogramming setup were that the model
\begin{inparaenum}[(i)]
    \item   offers state-of-the-art performance in audio classification,
    \item   is trained on a comparable but different task,
    \item   is preferably an attention-based model to make it better suited to work on the weakly labeled OpenMIC dataset (see below),
    \item   is of sufficiently high complexity to learn a powerful representation, and
    \item   has been trained on a large number of data points.
\end{inparaenum}

Given these criteria, the Audio Spectrogram Transformer (AST) \cite{Gong2021ASTAS,Gong2021SSASTSA} was selected. AST is a convolution-free, purely attention-based model with an audio spectrogram input, achieving state-of-the-art results on AudioSet \cite{7952261}, ESC-50 \cite{piczak2015dataset}, and Speech Commands V2 \cite{Warden2018SpeechCA}. Choosing the AST model trained on AudioSet provides us with a pre-trained model that should be suitable for music audio. The input audio is pre-processed into 128-dimensional Log-Mel spectrogram features computed with a \unit[25]{ms} von-Hann window every \unit[10]{ms}. The spectrogram is split into a sequence of $16\times16$ patches with overlap, and then linearly projected and added to a learnable positional embedding as the input of the transformer encoder. For the AST pre-trained on AudioSet, the number of output classes is 527.

\subsection{Reprogramming}
\label{repr}
The overall structure follows the flow-chart presented in Figure~\ref{fig:repr} with the reprogramming stage, the pre-trained model, and the output label mapping stage. 
To explore the potential and performance of reprogramming, we investigate various forms of input and output reprogramming in this work. 

\subsubsection{Input Reprogramming} 
\label{inre} 
 The input reprogramming step aims to find a trainable input modification that can be applied to inputs universally to transform them into an input representation useful for the pre-trained model. Previously proposed methods include the superposition of noise to the input also known as adversarial reprogramming \cite{elsayed2018adversarial,tsai2020transfer}. We investigate this approach, but we also propose a novel method to extend this simple superposition by transforming the input spectrogram by means of a neural network. In this way, a well-crafted perturbation of the input might improve ``compatibility'' with the pre-trained model. 

\textit{Noise Reprogramming}:
Previous reprogramming methods add a learnable noise component to input to translate the target data to the source domain of the pre-trained model. This is similar to what many approaches to adversarial attacks do\cite{elsayed2018adversarial,tsai2020transfer}. Unlike previous methods, we choose to add the noise to the spectrogram and not the time domain signal. We hypothesize that a spectrogram is a more fitting representation, because of the higher complexity of music audio compared to other signals reprogramming has been applied to. The operation can be formulated as $\hat{X} = X + N.$
$X$ is the input spectrogram with size $T \times F$ where $T$ is the number of time bins, and $F$ is the number of mel-band filters. $\hat{X}$ is the input of the pre-trained model. The learnable noise $N$ is universal to all target data and independent of $X$. The dimension of $N$ is identical to $X$, the size of the input spectrogram.

\textit{CNN Reprogramming}\label{cnn}:
The application of CNN to audio spectrograms is considered a standard baseline in many audio classification tasks. Therefore, we propose to use trainable CNN layers as an input transformation of the input data. These layers replace the noise superposition as input processing.
The transformation can be formally described by $\hat{X} = F(X),$
in which $F$ represents the CNN consisting of two 2D convolutional layers with a receptive field of $3 \times 3$, a stride of $1 \times 1$, and a padding size of $1 \times 1$. 
Note that in this special case, no max-pooling is applied as the input dimension matches the input dimension of the pre-trained model. The difference between Noise Reprogramming and CNN Reprogramming is that CNN layers apply learnable transformation to the input itself instead of simply adding learnable noise that is independent of the input. Due to the idea of weight sharing in CNN, the training parameters needed are the CNN kernels; hence, the amount of parameters is considerable less than Noise Reprogramming. 

\begin{figure}
\vspace{-8mm}
\centering
\includegraphics[width=0.85\textwidth]{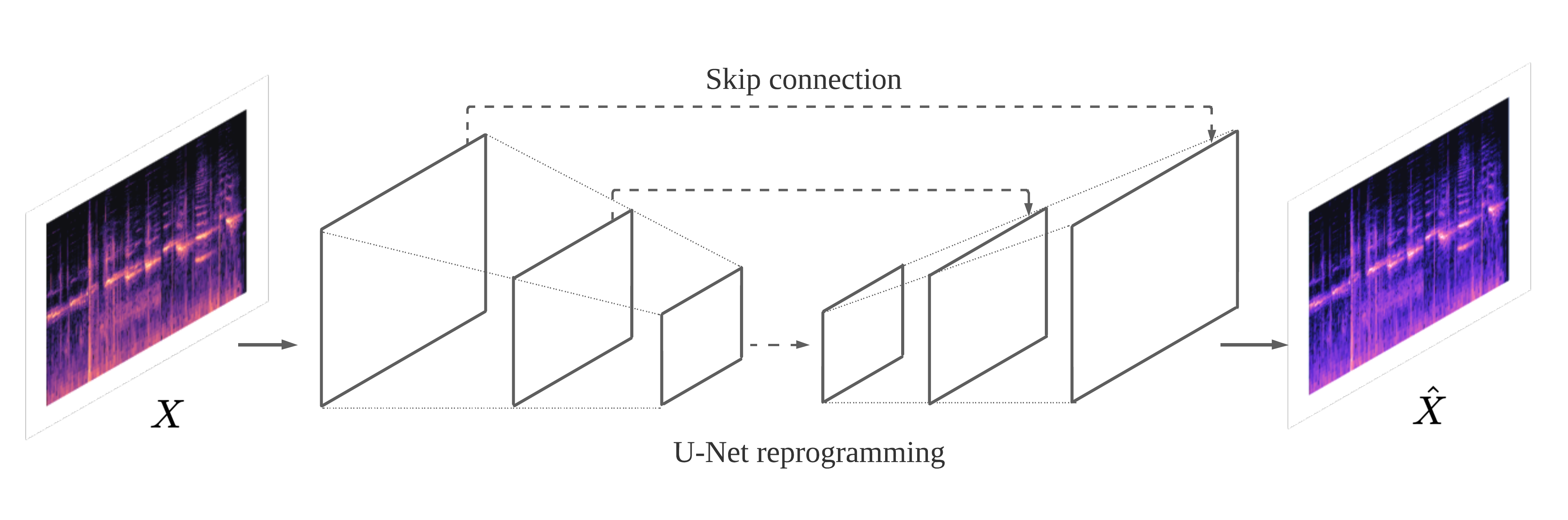}
\vspace{-3mm}
\caption{U-Net structure for input reprogramming.}
\label{unet}
\vspace{-7mm}
\end{figure}

\textit{U-Net Reprogramming}:
The idea behind using CNN Reprogramming is that the input audio spectrogram can be transformed into a suitable and compatible spectrogram for the pre-trained model. In order to provide more flexibility for the reprogramming learning, it is fair to consider the features at different time resolutions. 
Therefore, in addition to CNN reprogramming, we further propose U-Net reprogramming for MIC. U-Net is a CNN structure first developed for biomedical image segmentation \cite{Ronneberger2015UNetCN} and has become popular in both speech and music separation tasks. The architecture consists of a contraction path to capture context and a symmetric expansion path to reconstruct the extracted features back to input resolution \cite{Ronneberger2015UNetCN}. With convolutional layers and skip connections, U-Net is able to represent the input in both high-level features on coarser time-frequency scales and detailed features in deep CNN layers. These features are combined using bilinear upsampling blocks, yielding multi-scale features with both high-level and deep representations \cite{Jansson2017SingingVS}. 
The U-Net's success in music source separation tasks might also imply that U-Net processing on a spectrogram is effective to differentiate instrument content. To the best of our knowledge, this is the first time a U-Net structure has been proposed for reprogramming. 
The proposed U-Net structure for reprogramming is shown in Figure~\ref{unet}. The formulation is identical to the CNN reprogramming mentioned above as the transform function is applied to the input. It consists of three convolutional layers in both contraction path and expansion path, and each convolutional layer is followed by a batch normalization layer and a ReLU activation. A $2 \times 2$ max-pooling is applied for the CNN layer in the contraction path while upsampling is applied for that in expansion path.

\subsubsection{Output Reprogramming}
The output categories of the pre-trained system obviously do not match the music instrument classes to be classified. Therefore, the outputs of the pre-trained model have to be mapped to the target labels. For example, Yang et al.\ propose to use a many-to-one label mapping \cite{tsai2020transfer,yang2021voice2series}. 
For each target label, its class prediction will be the averaged class predictions over the set of source labels assigned to it. We investigate this approach, but also propose a new output mapping utilizing fully-connected (FC) layers to fit the targets. Here, the output probabilities are mapped to the target labels by the FC layer (FCL) and a sigmoid activation function. The mapping can thus be learned during the training phase. Note that the original last layer activation in AST is removed in our model.

\section{Experimental Setup}\label{sec:eval}
To evaluate the impact of different input and output reprogramming options, the results for the following systems will be reported:
\begin{compactitem}
    \item  \textit{Noise Reprogramming (AST-NRP)}: pre-trained model with added noise at the input and with output FCL label mapping,
    \item  \textit{CNN Reprogramming (AST-CNNRP)}: pre-trained model with CNN input processing and with output FCL label mapping, and
    \item  \textit{U-Net Reprogramming (AST-URP)}: pre-trained model with U-Net input processing and with output FCL label mapping.
\end{compactitem}
In addition to the three methods introduced above (AST-NRP, AST-CNNRP, AST-URP), we also add the following systems for comparison:
\begin{inparaenum}[(i)]
    \item the pre-trained AST (AST-BS) without input reprogramming as a baseline to evaluate the power of the AST representation for MIC,
    \item a CNN baseline (CNN-BS) with roughly the same number of training parameters as our proposed input transformation methods AST-CNNRP and AST-URP,
    \item a transfer learning approach by fine-tuning the AST system with the target data (AST-TL),
    \item the previous state-of-the-art Mean Teacher (MT) model by Gururani and Lerch \cite{9666061}, and
    \item the Random Forest (RF) baseline released with the OpenMic dataset \cite{humphrey2018openmic}.
\end{inparaenum}

The implementation of the proposed methods is publicly available.\footnote{\href{https://github.com/hchen605/ast_inst_cls}{github.com/hchen605/ast\_inst\_cls}, last accessed: Nov 9, 2022}

\subsection{Dataset}
\begin{figure}[t]
\centering
\includegraphics[width=\textwidth]{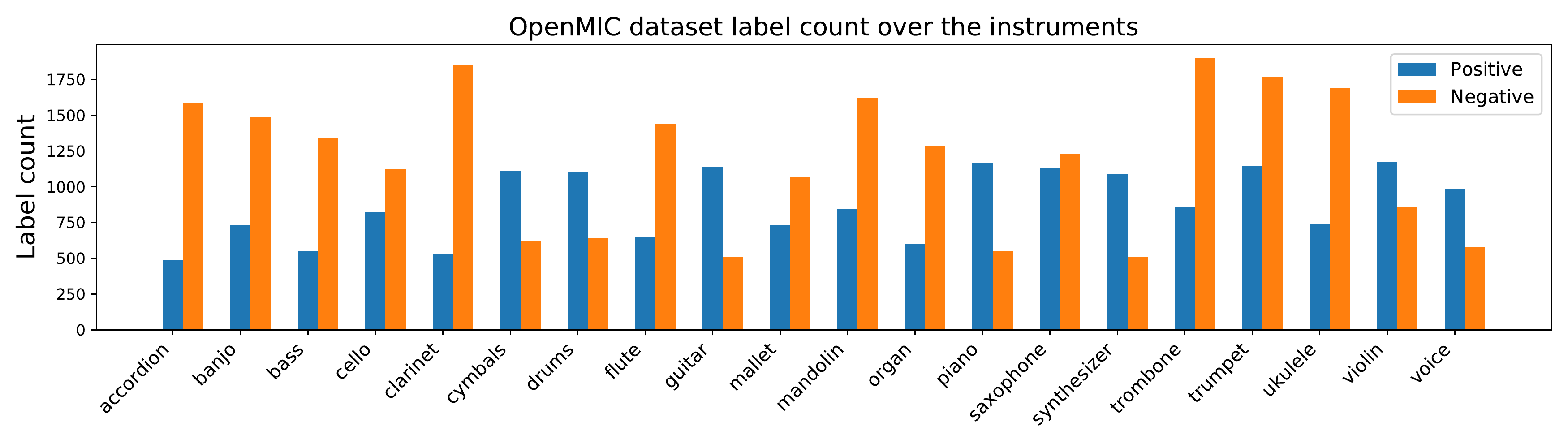}

\caption{OpenMIC: positive vs.\ negative labels.}
\label{fig:openmiclabel}
\vspace{-7mm}
\end{figure}

{We use the OpenMIC dataset for the experiments in this paper.}
OpenMIC is the first open multi-instrument music dataset with a comparably diverse set of musical instruments and genres, addressing issues in other previously existing datasets for MIC \cite{humphrey2018openmic}. It consists of 20,000 audio clips, each of \unit[10]{s} length. Every clip is labeled with the presence (positive) or absence (negative) of least one of 20 musical instruments, and each instrument class has at least 500 confirmed positives and at least totally 1500 confirmed labels. Note that if the dataset were fully labeled, it would come with $\unit[20000]{clips} \cdot\unit[20]{instruments} = \unit[400000]{labels}$, however, the actual number of labels in the dataset is $41,268$, meaning that approximately 90\% of the labels are missing. Moreover, each \unit[10]{s} clip has instrument presence or absence tags without specifying onset and offset times, also referred to as weak labels. Therefore, models cannot be trained using fine-grained instrument activity annotation. Figure~\ref{fig:openmiclabel} visualizes the overall label distribution of OpenMIC. We can observe the unbalanced nature; commonly seen instruments, such as piano, voice, and violin, generally have more positive labels than the others.  Another possible reason that we can see more positive labels for these common instruments might be that crowd-sourced annotators were more familiar with these common timbres. 

The models investigated in this study are trained to identify the presence or absence of musical instruments in OpenMIC dataset. The publicly available data splits are used for training and testing: approx.\ 25\% of data are used for testing, and 15\% of the training data is sampled randomly to form the validation set.

\subsection{Training Procedure and Evaluation Metrics}


The reprogramming model is trained with a batch size of 8, and the Adam optimizer with binary cross-entropy loss is used. We apply an initial learning rate of 5e-5 for the first 10 epochs, and then the learning rate is cut into half every 5 epochs until reaching 50 epochs.


We are interested in both the classification performance and the model complexity of each investigated system. To be comparable to previously reported results on the MIC task, the macro F1 score, calculated by purely averaging the per-instrument F1 scores and ignoring the weight of per-class data amount, is reported. The final results for each setup will be reported by averaging the macro F1 score of 10 experiments.
In addition to this classification metric, the number of training parameters of each of the evaluated systems will be reported. 

\subsection{Results and Discussion}
\label{exp}


\subsubsection{Input Reprogramming}

\begin{figure}[t]
\centering
\includegraphics[width=0.85\textwidth]{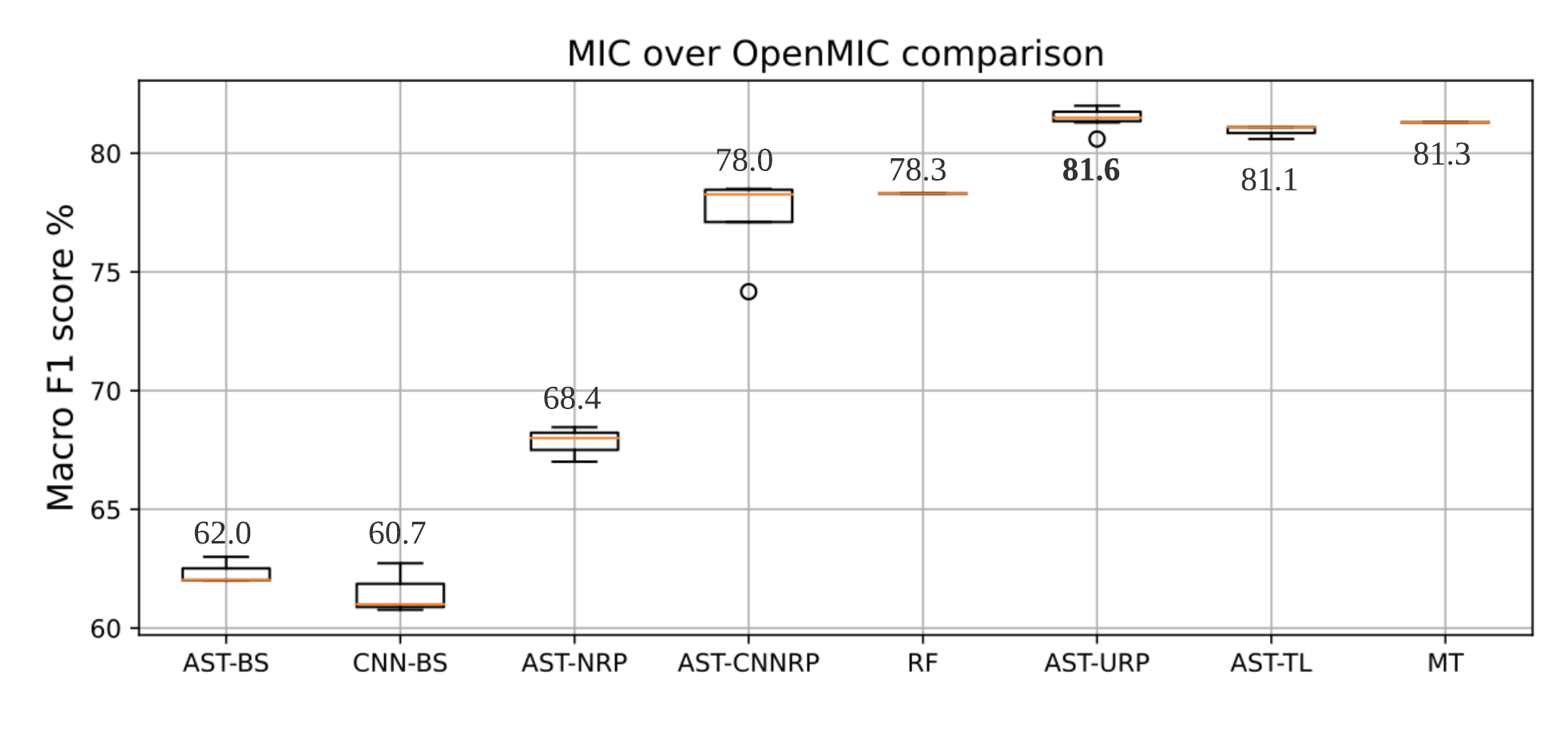}
\caption{Macro F1 scores of the evaluated methods.}
\label{repr1}
\vspace{-5mm}
\end{figure}

The overall average macro F1 scores are shown in Figure~\ref{repr1}. 
We note that the previous state-of-the-art MT \cite{9666061} is reported with the result 81.3\% as shown in the paper, and the RF baseline \cite{humphrey2018openmic} is reported with the result 78.3\% calculated from the released Python Notebook.
We can make the following observations.
First, simply using the pre-trained AST without input processing does not lead to convincing results, although at around 62\% the result is considerably higher than guessing (50\%). The performance of the un-tuned system roughly matches the performance of the simple CNN-BS trained on data for the task, indicating that AST has learned a powerful and useful representation.
\begin{table}[b]
\caption{Comparison of model training parameters. }
\centering
\label{complexity_2}
\begin{tabular*}{\columnwidth}{l @{\extracolsep{\fill}} |cccccccc}
Method & AST-BS  &  CNN-BS  & AST-TL & AST-NRP & AST-CNNRP & AST-URP & MT\\[1mm]
\hline\hline
\#Param.(M) & 0.017 & 0.017 & 87.873 & 0.148 & 0.017 & 0.018 & 0.111
\end{tabular*}
\vspace{-5mm}
\end{table}
Second, a trained noise signal that is simply added to the input can improve this baseline performance by more than 6\%, but the resulting system remains far from the performance of state-of-the-art systems. This implies that while traditional reprogramming approaches can work to a certain degree, the complexity and variability of music signals requires an input-adaptive transform as opposed to the addition a constant signal.
Third, the results for AST-TL show the effectiveness of transfer learning showing performance roughly on par with the state-of-the-art. This result emphasizes that transfer learning is a powerful tool with competitive results for weakly-labeled data.
Fourth, both the AST-CNNRP and AST-URP pre-processing steps dramatically improve classification performance over the AST-NRP with the U-Net-based reprogramming performing about 4\% better than the CNN. We also note that transfer learning with fine-tuning results in only limited improvement over the reprogramming methods, implying the AST representation without fine-tuning is suitable for this task.
Fifth, AST-URP slightly outperforms all presented system and even the state-of-the-art MT system.
\begin{figure*}
\centering
\includegraphics[width=0.9\textwidth]{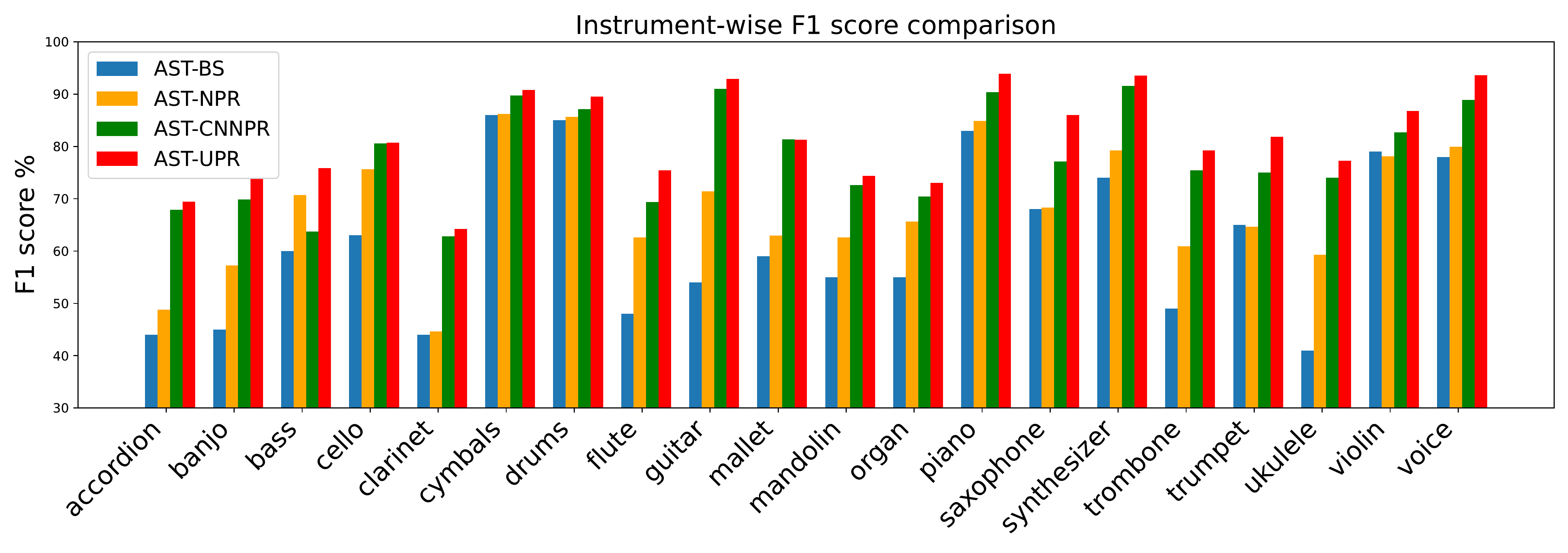}
\caption{F1 score comparison between the reprogramming methods.}
\label{cmp2}
\vspace{-7mm}
\end{figure*}
Diving deeper into the detailed results, Figure~\ref{cmp2} displays the instrument-wise F1 scores. We can observe a lot of variation in classification performance over instruments. These variations are related to the amount of positive labels as shown in Figure~\ref{fig:openmiclabel}, a clear indicator that the number of (positive) labels directly influences the classification performance. In fact, the correlation coefficient of the amount of instrument-wise positive labels and AST-BS F1 score is 0.75, showing the high correlation. 
We can see that the U-Net reprogramming model AST-URP outperforms all other models for all instruments, and we observe considerable performance gains especially for the instrument classes with few positive labels. For example, accordion has the lowest number of positive labels with around 500 and the AST-URP improvement is over 25\%. This demonstrates the capability of the proposed reprogramming method to solve the data scarcity issue by leveraging powerful other representations. 

The results support our assumption that an input transformation utilizing both high-level and low-level features benefits reprogramming. Overall, we can see that both AST-CNNRP and AST-URP are effective approaches to adapt a pre-trained model to a new task. It encourages future research on reprogramming pre-trained models with other methods of transformation.
\subsubsection{Complexity Analysis}
One of the main advantages of reprogramming is reduced training complexity as the pre-trained model remains unchanged. As an indicator of model complexity, Table~\ref{complexity_2} reports the number of training parameters along with the F1 score previously visualized in Figure~\ref{repr1}. In terms of complexity, it is clear that the training complexity of reprogramming is very low.
We can observe that MT, the previous state of the art system using a Mean Teacher approach with consistency loss has a higher complexity (one order of magnitude) than either our CNN-based approach or the high-performing U-Net approach, despite them using the VGGish representation as input to their system. Compared to the number of AST training parameters, the proposed AST-URP model has only a fraction of training parameters (0.02\%) but is still the best-performing system.



\section{Conclusion}
\label{sec:conclusion}
In this work, we propose to apply model reprogramming to the task of music instrument classification. We extend the existing reprogramming approaches by utilizing a novel U-Net-based input reprogramming method. By leveraging the power of pre-trained audio spectrogram transformer model, we show that our model can achieve state-of-the-art performance at a fraction of the training complexity of other models. We provide detailed results on the impact of different input and output reprogramming approaches. 
Without applying data augmentation and advanced model architectures like semi-supervised MT learning, reprogramming a pre-trained model is a low-complexity approach to achieving state-of-the-art performance. Although the workload during the inference stage remains unchanged, the low training complexity of this new transfer learning approach in combination with our promising results opens up a multitude of possible use cases in tasks in MIR and other fields with insufficient data. 


In future work, we plan to explore variations of the input reprogramming stages beyond CNN or U-Net structures. Furthermore, we plan to test the reprogramming algorithm with various other pre-trained models from other audio and non-audio tasks. We believe that reprogramming is a promising approach with the potential to be used in many other MIR tasks. 


%
%
%
\bibliographystyle{splncs04}
\bibliography{mmm}

\begin{thebibliography}{10}
\providecommand{\url}[1]{\texttt{#1}}
\providecommand{\urlprefix}{URL }
\providecommand{\doi}[1]{https://doi.org/#1}

\bibitem{1661252}
Benetos, E., Kotti, M., Kotropoulos, C.: Musical instrument classification
  using non-negative matrix factorization algorithms and subset feature
  selection. In: ICASSP (2006)

\bibitem{Biggio2018WildPT}
Biggio, B., Roli, F.: Wild patterns: Ten years after the rise of adversarial
  machine learning. Pattern Recognit.  \textbf{84},  317--331 (2018)

\bibitem{Bittner2014MedleyDBAM}
Bittner, R.M., Salamon, J., Tierney, M., Mauch, M., Cannam, C., Bello, J.P.:
  Medleydb: A multitrack dataset for annotation-intensive mir research. In:
  ISMIR (2014)

\bibitem{bosch2012comparison}
Bosch, J.J., Janer, J., Fuhrmann, F., Herrera, P.: A comparison of sound
  segregation techniques for predominant instrument recognition in musical
  audio signals. In: ISMIR (2012)

\bibitem{Choi2017TransferLF}
Choi, K., Fazekas, G., Sandler, M.B., Cho, K.: Transfer learning for music
  classification and regression tasks. In: ISMIR (2017)

\bibitem{dau2019ucr}
Dau, H.A., Bagnall, A., Kamgar, K., Yeh, C.C.M., Zhu, Y., Gharghabi, S.,
  Ratanamahatana, C.A., Keogh, E.: The {UCR} time series archive. IEEE/CAA
  Journal of Automatica Sinica  \textbf{6}(6),  1293--1305 (2019)

\bibitem{Defferrard2017FMAAD}
Defferrard, M., Benzi, K., Vandergheynst, P., Bresson, X.: {FMA}: A dataset for
  music analysis. In: ISMIR (2017)

\bibitem{elsayed2018adversarial}
Elsayed, G.F., Goodfellow, I., Sohl-Dickstein, J.: Adversarial reprogramming of
  neural networks. In: ICLR (2019)

\bibitem{859069}
Eronen, A., Klapuri, A.: Musical instrument recognition using cepstral
  coefficients and temporal features. In: ICASSP (2000)

\bibitem{1661401}
Essid, S., Richard, G., David, B.: Hierarchical classification of musical
  instruments on solo recordings. In: ICASSP (2006)

\bibitem{7952261}
Gemmeke, J.F., Ellis, D.P.W., Freedman, D., Jansen, A., Lawrence, W., Moore,
  R.C., Plakal, M., Ritter, M.: Audio set: An ontology and human-labeled
  dataset for audio events. In: ICASSP (2017)

\bibitem{Gong2021ASTAS}
Gong, Y., Chung, Y.A., Glass, J.R.: {AST}: Audio spectrogram transformer. In:
  Interspeech (2021)

\bibitem{Gong2021SSASTSA}
Gong, Y., Lai, C.I., Chung, Y.A., Glass, J.R.: {SSAST}: Self-supervised audio
  spectrogram transformer. In: AAAI (2021)

\bibitem{gururani2017mixing}
Gururani, S., Lerch, A.: Mixing secrets: A multi-track dataset for instrument
  recognition in polyphonic music. In: ISMIR (2017)

\bibitem{9666061}
Gururani, S., Lerch, A.: Semi-supervised audio classification with partially
  labeled data. In: IEEE ISM (2021)

\bibitem{DBLP:conf/ismir/GururaniSL19}
Gururani, S., Sharma, M., Lerch, A.: An attention mechanism for musical
  instrument recognition. In: ISMIR (2019)

\bibitem{hambardzumyan2021warp}
Hambardzumyan, K., Khachatrian, H., May, J.: {WARP}: {W}ord-level {A}dversarial
  {R}e{P}rogramming. In: IJCNLP (2021)

\bibitem{he2016deep}
He, K., Zhang, X., Ren, S., Sun, J.: Deep residual learning for image
  recognition. In: CVPR (2016)

\bibitem{hershey2017cnn}
Hershey, S., Chaudhuri, S., Ellis, D.P., Gemmeke, J.F., Jansen, A., Moore,
  R.C., Plakal, M., Platt, D., Saurous, R.A., Seybold, B., et~al.: {CNN
  Architectures for Large-scale Audio Classification}. In: ICASSP (2017)

\bibitem{Howard2017MobileNetsEC}
Howard, A.G., Zhu, M., Chen, B., Kalenichenko, D., Wang, W., Weyand, T.,
  Andreetto, M., Adam, H.: Mobilenets: Efficient convolutional neural networks
  for mobile vision applications. CVPR  (2017)

\bibitem{humphrey2018openmic}
Humphrey, E., Durand, S., McFee, B.: Openmic-2018: An open data-set for
  multiple instrument recognition. In: ISMIR (2018)

\bibitem{DBLP:conf/ismir/HungY18}
Hung, Y., Yang, Y.: Frame-level instrument recognition by timbre and pitch. In:
  ISMIR (2018)

\bibitem{Jansson2017SingingVS}
Jansson, A., Humphrey, E.J., Montecchio, N., Bittner, R.M., Kumar, A., Weyde,
  T.: Singing voice separation with deep u-net convolutional networks. In:
  ISMIR (2017)

\bibitem{9229505}
Kong, Q., Cao, Y., Iqbal, T., Wang, Y., Wang, W., Plumbley, M.D.: {PANN}s:
  Large-scale pretrained audio neural networks for audio pattern recognition.
  IEEE/ACM Transactions on Audio, Speech, and Language Processing  (2020)

\bibitem{Li2015AutomaticIR}
Li, P.Q., Qian, J., Wang, T.: Automatic instrument recognition in polyphonic
  music using convolutional neural networks. CoRR  \textbf{abs/1511.05520}
  (2015)

\bibitem{Lostanlen2016DeepCN}
Lostanlen, V., Cella, C.E.: Deep convolutional networks on the pitch spiral for
  music instrument recognition. In: ISMIR (2016)

\bibitem{8256990}
Nagawade, M.S., Ratnaparkhe, V.R.: Musical instrument identification using
  {MFCC}. In: RTEIC) (2017)

\bibitem{neekhara2018adversarial}
Neekhara, P., Hussain, S., Dubnov, S., Koushanfar, F.: Adversarial
  reprogramming of text classification neural networks. In: EMNLP-IJCNLP (2019)

\bibitem{piczak2015dataset}
Piczak, K.J.: {ESC}: {Dataset} for {Environmental Sound Classification}. In:
  ACM MM. pp. 1015--1018. {ACM} (2015)

\bibitem{pons2019musicnn}
Pons, J., Serra, X.: Musicnn: Pre-trained convolutional neural networks for
  music audio tagging. ISMIR  (2019)

\bibitem{Ronneberger2015UNetCN}
Ronneberger, O., Fischer, P., Brox, T.: {U-Net}: Convolutional networks for
  biomedical image segmentation. In: MICCAI (2015)

\bibitem{tan2018survey}
Tan, C., Sun, F., Kong, T., Zhang, W., Yang, C., Liu, C.: A survey on deep
  transfer learning. In: ICANN. Springer (2018)

\bibitem{tsai2020transfer}
Tsai, Y.Y., Chen, P.Y., Ho, T.Y.: Transfer learning without knowing:
  Reprogramming black-box machine learning models with scarce data and limited
  resources. In: ICML. PMLR (2020)

\bibitem{Warden2018SpeechCA}
Warden, P.: Speech commands: A dataset for limited-vocabulary speech
  recognition. CoRR  \textbf{abs/1804.03209} (2018)

\bibitem{yang2021voice2series}
Yang, C.H.H., Tsai, Y.Y., Chen, P.Y.: Voice2series: Reprogramming acoustic
  models for time series classification. In: ICML. PMLR (2021)

\end{thebibliography}

\end{document}